\documentclass[12pt]{article}
\usepackage{amsmath}
\usepackage{amssymb}
\usepackage{graphics}
\usepackage{epsf,graphicx,rotating,color}
\usepackage{bm,bbm}
\usepackage{amsfonts}

\def\r{\bf r}
\def\R{\bf R}

\begin{document}
\title{{\bf   Damping-Antidamping Effect on Comets Motion}}
\author{ G.V.  L\'opez\footnote{gulopez@udgserv.cencar.udg.mx}~ and E. M. Ju\'arez\\\\Departamento de F\'{\i}sica de la Universidad de Guadalajara,\\
Blvd. Marcelino Garc\'{\i}a Barrag\'an 1421, esq. Calzada Ol\'{\i}mpica,\\
44430 Guadalajara, Jalisco, M\'exico03.67.Lx,\  03.67.Hr,\  03.67.-a,\  03.65w}
\date{September, 2013}
\maketitle

\begin{abstract}
We make an observation about Galilean transformation on a 1-D mass variable systems which  leads  us to the right way to deal with mass variable systems. Then using this observation, we study two-bodies gravitational problem  where the mass of one of the bodies varies and suffers a damping-antidamping effect due to star wind during its motion. For this system, a constant of motion, a Lagrangian and a Hamiltonian are given for the radial motion, and the period of the body is studied using the constant of motion of the system.  Our theoretical results are applied to  Halley's comet.
\end{abstract}
\newpage
\section{\bf Introduction}
\noindent
There is not doubt that mass variable systems have been relevant since the foundation of the classical mechanics and
modern physics (L\'opez et al 2004). These type of systems  have been known as  Gylden-Meshcherskii problems (Gylden 1884;  Meshcherskii 1893, 1902;  Lovett 1902; Jeans 1924;  Berkovich 1981; Bekov 1989; Prieto and Docobo 1997), and  among these type of systems one could mention: the motion of rockets  (Sommerfeld 1964), the kinetic theory of dusty plasma (Zagorodny et al 2000), propagation of electromagnetic waves in a dispersive-nonlinear media (Serimaa et al 1986), neutrinos mass oscillations (Bethe 1986;  Commins and Bucksbaum 1983), black holes formation (Helhl et al 1998), and comets interacting with solar wind (Daly  1989). This last system belong to the  so called "gravitational two-bodies problem" which is one of the most studied and well known system in classical
mechanics (Goldstein 1950). In this type of system, one assumes normally that the masses of the bodies are fixed and unchanged during the dynamical motion.  However,when one is dealing with comets, beside to consider its mass variation due to the interaction with the solar wind, one would like to have an estimation of the the effect of the solar wind pressure on the comet motion. This pressure may produces a dissipative-antidissipative effect on its motion. The dissipation effect must be felt by the comet when this one is approaching to the sun (or star), and the antidissipation effect must be felt by
the comet when this one is moving away from the sun. To deal with these type of mass variation problem, it has been proposed that the Newton equation must be modified (Sommerfeld 1964; Plastino and Muzzio 1992) since the system becomes non invariant under change of inertial systems (Galileo transformation).\\Ê\\
 In this paper, we will make first an observation about this statement which indicates the such a proposed modification of Newton's equation has some problems and rather  the use of the original Newton equation is the right approach to deal with mass variation systems, which it was used in  previous paper (L\'opez 2007) to  study  two-bodies gravitational problem with mass variation in one of them, where we were interested in the difference of the  trajectories in the spaces ($x,v$) and ($x,p$). As a consequence, there is an indication that mass variation problems must be dealt as non invariant under Galilean transformation. 
Second, we study the two-bodies gravitational problem taking into consideration the mass variation of one of them  and its damping-antidamping effect due to the solar wind. The mass of the other body is assumed big and fixed , and the reference system of motion  is chosen just in this body. In addition, we will assume that the mass lost is expelled from the body radially to its motion. Doing this, the three-dimensional two-bodies problem is
reduced to a one-dimensional problem. Then, a constant of motion, the Lagrangian, and the Hamiltonian are deduced for  this one-dimensional problem, where a radial dissipative-antidissipative force proportional to the velocity square is
chosen. A model for the mass variation is given, and the damping-antidamping effect is studied on the period of the trajectories, the trajectories themselves, and the aphelion distance of a comet.   We use the parameters associated to comet Halley to illustrate the application of our results.\\Ê\\ 
\noindent
\section{\bf Mass variation problem and Galileo transformation.}
To simplify our discussion and  without  losing  generality, we will restrict myself to one degree of freedom,.  Newton equation of motion  is given by
\begin{equation}\label{g1}
\frac{d}{dt}\bigl(m(t)v\bigr)=F(x,v,t) ,
\end{equation}
where $m(t){\bf v}$ is the quantity of movement, $F$ is the total external force acting on the object, $m(t)$ and $v=dx/dt$ are its time depending mass and velocity of the body (motion of the mass lost is not considered). Galileo transformations to another inertial frame ($S'$)  which is moving with a constant velocity $u$ respect our original frame $S$ are defined as
\begin{subequations}\label{g2}
\begin{equation}
x'=x-ut
\end{equation}
\begin{equation}
t'=t
\end{equation}
\end{subequations}
which implies the following relation between the velocity seen in the reference system $S$, $v$,  and the velocity seen in the reference system $S'$, $v'$, 
\begin{equation}\label{g2}
 v'=v-u .
 \end{equation}
Multiplying the last term  by $m(t')$ and making the differentiation with respect to $t'$, one gets
\begin{equation}\label{g3}
\frac{d}{dt'}\bigl(m(t')v'\bigr)=F'(x',v',t')\ ,
\end{equation}
where $ F'$ is given by
\begin{equation}\label{g4}
F'(x',v',t')=F(x'+ut',v'+u,t')-u \frac{dm(t')}{dt'}.
\end{equation}
 Therefore, Eq. \ref{g1} and Eq. \ref{g3} have the same form but the force is different since in addition to the transformed force term $F(x'+ut',v'+u,t')$, one has the term $u dm(t')/dt'$. This non invariant form of the force under Galilean transformation has lead to propose  (Sommerfeld 1964; Plastino and Muzzio 1992) that Newton equation (\ref{g1})  to modify Newton's equation of motion for mass variation objects,  to keep the principle of invariance of equation under Galilean transformations, of the form 
\begin{equation}\label{g5}
m(t)\frac{dv}{dt}=F(x,v,t)+w\frac{dm(t)}{dt}\ ,
\end{equation}
where $w$ is the relative velocity of the escaping mass with respect the center of mass of the object. When one does a Galilean transformation on this equation, one gets 
\begin{equation}\label{g6}
m(t')\frac{dv'}{dt'}=F'(x',v',t') ,
\end{equation}
where $F'$ is given by
\begin{equation}\label{g7}
F'(x',v',t')=F(x'+ut',v'+u,t')+w\frac{dm(t')}{dt'},
\end{equation}
which has the same form as Eq. \ref{g5}. However, assume for the moment that $w=constant$  and $F=0$. So, from Eq. \ref{g5},  it follows that
\begin{equation}\label{g8}
v(t)=v_0+\ln\left(\frac{m(t)}{m_0}\right)^w ,
\end{equation}
where $m_0=m(0)$. In this way, if we have a mass variation of the for $m(t)=m_0e^{-\alpha t}$ (for example), one would have a velocity behavior like
\begin{equation}\label{g9}
v(t)=v_0-w\alpha t, 
\end{equation}
which is not acceptable since one can have $v>0$, $v=0$ and $v<0$ depending on the value $w\alpha t$. Even more, since for $F=0$, the equation resulting in the reference system $S'$ is the same, i.e. in $S'$ one gets the same type of solution, 
\begin{equation}\label{g10}
v'(t')=v_0+\ln\left(\frac{m(t')}{m_0}\right)
\end{equation}
which is independent on the relative motion of the reference frames, and this must  not be possible due to  relation (\ref{g2}).  \\Ê\\
In addition, it worths to mention that special theory of relativity can be seen as the motion of mass variation problem, where the mass depends on the velocity of the particle of the form $m(v)=m_0(1-v^2/c2)^{-1/2}$, with $c$ being the speed of light. This system is obviously not invariant under Galilean transformation,  and given the force, Newton's equation motion is always kept  in the same form to solve a relativistic problem, $d\bigl(m(v)v\bigr)/dt=F(x,v,t)$, (C. M{\o}ller 1952,  L\'opez et al 2004). 
\section{\bf  Mass variation and equations of motion.}
\noindent
Having explained and clarify the problem of mass variation (Spivak 2010), 
Newton's equations of motion for two bodies interacting gravitationally, seen from arbitrary
inertial reference system, and with radial dissipative-antidissipative force acting in one of them are given by
\begin{subequations}
\begin{equation}\label{em1a}
{d\over dt}\left(m_1{d{\r_1}\over dt}\right)=-{Gm_1m_2\over
|{\r_1}-{\r_2}|^3}(\r_1-\r_2)
\end{equation}
 and
 \begin{equation}\label{em1b}
{d\over dt}\left(m_2{d{\r_2}\over dt}\right)=-{Gm_1m_2\over |{\r_2}-{\r_1}|^3}(\r_2-\r_1)
-{\gamma\over|\r_1-\r_2|}\left[{d|\r_1-\r_2|\over dt}\right]^2(\r_2-r_1)\ ,
\end{equation}
\end{subequations}
where $m_1$ and $m_2$ are the masses of the two bodies, ${\r_1}=(x_1,y_1,z_1)$ and \hfil\break
${\r_2}=(x_2,y_2,z_2)$ are their vectors positions from the reference system, $G$ is
the gravitational constant ($G=6.67\times 10^{-11}m^3/Kg~s^2$), $\gamma$ is the nonnegative constant parameter of the
dissipative-antidissipative force, and 
\begin{equation*} 
{|\r_1-\r_2|=|\r_2-\r_1|}=\sqrt{(x_2-x_1)^2+(y_2-y_1)^2+(z_2-z_1)^2}
\end{equation*}
is the Euclidean distance between the two bodies. Note that if $\gamma> 0$ and \\ $d|\r_1-\r_2|/dt>0$ one has dissipation since the force acts against the motion of the body, and for $d|\r_1-\r_2|/dt<0$ one has anti-dissipation since the force pushes the body. If $\gamma<0$ this scheme  is reversed and  corresponds to our actual situation with the comet mass lost.\\\\
\noindent
It will be assumed the mass $m_1$ of the first body is constant and that the mass $m_2$ of the second body varies. Now, It is clear that the usual
relative, $\r$, and center of mass, $\R$, coordinates defined as
${\r}={\r_2}-{\r_1}$ and ${\R}=(m_1{\r_1}+m_2{\r_2})/( m_1+m_2)$ are not so good to describe the dynamics of this system.
However, one can consider the case for $m_1\gg m_2$ (which is the case star-comet), and consider
to put our reference system just on the first body ($\r_1=\vec 0$). In this case, Eq. (\ref{em1a}) and Eq. (\ref{em1b}) are reduced to the equation
\begin{equation}\label{em2}
m_2{d^2{\r}\over dt^2}=-{Gm_1m_2\over r^3}~{\r}-\dot{m}_2\dot{\r}-{\gamma}\left[{dr\over dt}\right]^2\hat {\r}\ ,
\end{equation}
where one has made the definition ${\r}={\r_2}=(x,y,z)$,  $r$ is its magnitude, $r=\sqrt{x^2+y^2+z^2}$, and $\hat{\r}={\r}/r$ is the unitary radial vector.  Using
spherical coordinates ($r,\theta,\varphi$),
\begin{equation}\label{em3}
x=r\sin\theta\cos\varphi\ ,\ \ y=r\sin\theta\sin\varphi\ ,\ \ z=r\cos\theta\ ,
\end{equation}
one obtains the following coupled equations
\begin{subequations}
\begin{equation}\label{em4a}
m_2(\ddot{r}-r\dot{\theta}^2-r\dot{\varphi}^2\sin^2\theta)=-{Gm_1m_2\over r^2}-\dot{m}_2\dot{r}-\gamma\dot{r}^2\
,\end{equation}
\begin{equation}\label{em4b}
m_2(2\dot{r}\dot{\theta}+r\ddot{\theta}-r\dot{\varphi}^2\sin\theta\cos\theta)=
-\dot{m}_2r\dot{\theta}\ ,\end{equation}
and
\begin{equation}\label{em4c}
m_2(2\dot{r}\dot{\varphi}\sin\theta+r\ddot{\varphi}\sin\theta+2r\dot{\varphi}\dot{\theta}\cos\theta)
=-\dot{m}_2r\dot{\varphi}\sin\theta\ .\end{equation}
\end{subequations}
Taking $\dot{\varphi}=0$  as solution of this last equation,  the resulting equations are
\begin{subequations}
\begin{equation}\label{em5a}
m_2(\ddot{r}-r\dot{\theta}^2)=-{Gm_1m_2\over r^2}-\dot{m}_2\dot{r}-\gamma\dot{r}^2\ ,
\end{equation}
and
\begin{equation}\label{em5b}
m_2(2\dot{r}\dot{\theta}+r\ddot{\theta})+\dot{m}_2r\dot{\theta}=0\ .
\end{equation}
\end{subequations}
From this last expression,   one  gets the following constant of motion (usual angular momentum of the system) 
\begin{equation}\label{em6a}
l_{\theta}=m_2r^2\dot{\theta}\ ,
\end{equation}
and with this constant of motion substituted in  Eq. (\ref{em5a}), one obtains the following one-dimensional equation of motion for the radial part 
\begin{equation}\label{em6}
{d^2r\over dt^2}=-{Gm_1\over r^2}-{\dot{m}_2\over m_2}\left({dr\over dt}\right)
-{\gamma\over m_2}\dot{r}^2+{l_{\theta}^2\over m_2^2r^3}\ .
\end{equation}
Now, let us assume that $m_2$ is a function of the distance between the first and the second body, $m_2=m_2(r)$.
Therefore, it follows that 
\begin{equation}\label{em7}
\dot{m}_2=m_2'\dot{r}\ ,
\end{equation}
where $m_2'$ is defined as $m_2'=dm_2/dr$. Thus, Eq. (\ref{em6}) is written as
\begin{equation}\label{em8}
{d^2r\over dt^2}=-{Gm_1\over r^2}+{l_{\theta}^2\over m_2^2 r^3}
-{{m}_2'+\gamma\over m_2}\left({dr\over dt}\right)^2\ ,
\end{equation} 
which, in turns, can be  written as the following autonomous dynamical system 
\begin{equation}\label{em9}
{dr\over dt}=v\ ;\quad\quad{dv\over dt}=-{Gm_1\over r^2}+{l_{\theta}^2\over m_2^2r^3}
-{{m}_2'+\gamma\over m_2}v^2\ .
\end{equation}
Note from this equation that $m_2'$ is always a non-positive function of $r$ since it represents the mass lost rate. On the other hand,  $\gamma$ is a negative parameter in our case. 
\section{\bf Constant of Motion, Lagrangian and Hamiltonian}
\noindent
A constant of motion for the dynamical system (\ref{em9}) is a function $K=K(r,v)$ which
satisfies the partial differential equation (L\'opez 1999)
\begin{equation}\label{cm1}
v{\partial K\over\partial r}+\left[{-Gm_1\over r^2}+{l_{\theta}^2\over m_2^2 r^3}
-{m_2'+\gamma\over m_2}v^2\right]{\partial K\over\partial v}=0\ .
\end{equation}
The general solution of this equation is given by (John 1974)
\begin{equation}\label{cm2}
K(x,v)=F(c(r,v))  ,
\end{equation}
where $F$ is an arbitrary function of the characteristic curve $c(r,v)$ which has the following expression
\begin{equation}\label{cm3}
c(r,v)=m_2^2(r)e^{2\gamma\lambda(r)}v^2+\int\left({2Gm_1\over r^2}-{2l_{\theta}^2\over m_2^2r^3}\right)m_2^2(r) 
e^{2\gamma\lambda(r)}dr ,
\end{equation}
and the function $\lambda(r)$ has been defined as
\begin{equation}\label{cm4}
\lambda(r)=\int{dr\over m_2(r)} .
\end{equation}
During a cycle of oscillation, the function $m_2(r)$ can be different for the comet approaching the sun and for the
comet moving away from the sun. Let us denote $m_{2+}(r)$ for the first case and $m_{2-}(r)$ for the second case.
Therefore, one has two cases to consider in Eq. (\ref{cm2}) which will denoted by ($\pm$). Now, if 
$m_{2\pm}^o$ denotes the mass at aphelium (+) or perielium (-) of the comet, $F(c)=c^{\pm}/2m_{2\pm}^o$ represents
the functionality in Eq. (\ref{cm2}) such that for $m_2$ constant and $\gamma$ equal zero, this constant of motion is the
usual gravitational energy. Thus, the constant of motion can be chosen as $K^{\pm}=c(r,v)/2m_{2\pm}^0$, that is, 
\begin{subequations}
\begin{equation}\label{cm5a}
K^{\pm}={m_{2\pm}^2(r)\over 2m_{2\pm}^o}e^{2\gamma\lambda_{\pm}(r)}v^2+V_{eff}^{\pm}(r)  ,\end{equation}
where the effective potential $V_{eff}$ has been defined as
\begin{equation}
V_{eff}^{\pm}(r)={Gm_1\over m_{2\pm}^o}\int{m_{2\pm}^2(r)e^{2\gamma\lambda_{\pm}(r)}dr\over r^2}
-{l_{\theta}^2\over m_{2\pm}^o}       \int{e^{2\gamma\lambda_{\pm}(r)}dr\over r^3}\end{equation}
\end{subequations}
This effective potential has an extreme at the point $r_*$ defined by the relation
\begin{equation}\label{cm6}
r_*m_2^2(r_*)={l_{\theta}^2\over Gm_1}\end{equation}
which is independent on the parameter $\gamma$ and depends on the behavior of $m_2(r)$. This extreme point is a minimum of the effective potential
since one has
\begin{equation}\label{cm7}
\left({d^2V_{eff}^{\pm}\over dr^2}\right)_{r=r_*}>0 .\end{equation}
Using the known expression (Kobussen 1979; Leubner 1981; L\'opez 1996) for the Lagrangian in terms of the constant of motion,
\begin{equation}\label{cm8}
L(r,v)=v\int{K(r,v)~dv\over v^2} ,\end{equation}
the Lagrangian, generalized linear momentum and the Hamiltonian are given by
\begin{equation}\label{cm9}
L^{\pm}={m_{2\pm}^2(r)\over 2m_{2\pm}^o}e^{2\gamma\lambda_{\pm}(r)}v^2-V_{eff}^{\pm}(r) ,\end{equation}
\begin{equation}\label{cm10}
p={m_{2\pm}^2(r)~v\over m_{2\pm}^o}e^{2\gamma\lambda_{\pm}(r)} ,\end{equation}
and
\begin{equation}\label{cm11}
H^{\pm}={m_{2\pm}^op^2\over 2m_{2\pm}^2(r)}e^{-2\gamma\lambda_{\pm}(r)}+V_{eff}^{\pm}(r)  .\end{equation}
The trajectories in the space ($x,v$) are determined by the constant of motion (\ref{cm5a}). Given the initial condition
($r_o,v_o$), the constant of motion has the specific value
\begin{equation}\label{cm12}
K^{\pm}_o={m_{2\pm}^2(r_o)\over 2m_{2\pm}^o}e^{2\gamma\lambda_{\pm}(r_o)}v^2_o+V_{eff}^{\pm}(r_o) ,
\end{equation}
and the trajectory in the space ($r,v$) is given by
\begin{equation}\label{cm13}
v=\pm\sqrt{2m_{2\pm}^o\over m_{2\pm}^2(r)}~e^{-\gamma\lambda_{\pm}(r)}\biggl[K_o^{\pm}-V_{eff}^{\pm}(r)\biggr]^{1/2}
 .\end{equation}
Note that one needs to specify $\dot{\theta}_o$ also to determine Eq. (\ref{em6a}). In addition, one normally wants to know
the trajectory in the real space, that is, the acknowledgment of $r=r(\theta)$. Since one has that
$v=dr/dt=(dr/d\theta)\dot\theta$ and Eqs. (\ref{em6a}) and (\ref{cm13}), it follows that
\begin{equation}\label{cm14}
\theta(r)=\theta_o+{l_{\theta}^2\over\sqrt{2m_{2\pm}^o}}\int_{r_o}^r{m_{2\pm}(r)e^{\gamma\lambda_{\pm}(r)}dr
\over r^2\sqrt{K_o^{\pm}-V_{eff}^{\pm}(r)}} .\end{equation}
The half-time period (going from aphelion to perihelion (+), or backward (-)) can be deduced from Eq. (\ref{cm13}) as
\begin{equation}\label{cm15}
T_{1/2}^{\pm}={1\over\sqrt{2m_{2\pm}^o}}\int_{r_1}^{r_2}{m_{2\pm}(r)e^{\gamma\lambda_{\pm}(r)}dr
\over \sqrt{K_o^{\pm}-V_{eff}^{\pm}(r)}} ,\end{equation}
where $r_1$ and $r_2$ are the two return points  resulting from the solution of the following equation
\begin{equation}\label{cm16}
V_{eff}^{\pm}(r_i)=K_o^{\pm}\quad i=1,2 .\end{equation}
On the other hand, the trajectory in the
space ($r,p$) is determine by the Hamiltonian (\ref{cm11}), and given the same initial conditions, the initial $p_o$ and
$H_o^{\pm}$ are obtained from Eqs. (\ref{cm11}) and  (\ref{cm10}). Thus, this trajectory is given by
\begin{equation}\label{cm17}
p=\pm\sqrt{2m_{2\pm}^2(r)\over m_{2\pm}^o}~e^{\gamma\lambda_{\pm}(r)}\biggl[H_o^{\pm}-V_{eff}^{\pm}(r)\biggr]^{1/2}
 .\end{equation}
It is clear just by looking the expressions (\ref{cm13}) and (\ref{cm17}) that the trajectories in the spaces ($r,v$) and ($r,p$) must be different due to complicated relation (\ref{cm10}) between $v$ and $p$ (L\'opez 2007).\\\\
\section{\bf  Mass-Variable Model and Results}
\noindent
As a possible application, consider that a comet looses material as a result of
the interaction with star wind in the following way (for one cycle of oscillation)
\begin{equation}\label{mv1}
m_{2\pm}(r)=\left\{\begin{array}{l l} 
m_{2-}(r_{2(i-1)})\biggl(1-e^{-\alpha r}\biggr)& incoming (+)\  v<0\\ \\
m_{2+}(r_{2i-1})-b\biggl(1-e^{-\alpha (r-r_{_{2i-1}})}\biggr)& outgoing (-)\  v>0
 \end{array}\right.
\end{equation}
where the parameters $b>0$ and $\alpha>0$ can be chosen to math  the mass loss rate in the incoming and outgoing cases. The
index "i" represent the ith-semi-cycle, being $r_{2(i-1)} $ and $r_{2i-1} $ the aphelion($r_a$) and perihelion($r_p$) points ($r_o$ is given by the initial conditions, and one has that $m_{2-}(r_o)=m_o$). For this case, the functions $\lambda_{+}(r)$ and
$\lambda_{-}(r)$ are given by
\begin{subequations}
\begin{equation}\label{mv2a}
\lambda_{+}(r)={1\over\alpha m_a}~\ln{\biggl(e^{\alpha r}-1\biggr)} ,\end{equation}
and
\begin{equation}\label{mv2b}
\lambda_{-}(r)={ -1\over \alpha(b-m_p)}
\Biggl[\alpha r+\ln{\left(m_p-b(1-e^{-\alpha(r-r_p)})\right)}\Biggr] .\end{equation}
\end{subequations}
where we have defined $m_a=m_2(r_a)$ and $m_p=m_2(r_p)$. Using the Taylor expansion, one gets
\begin{subequations}
\begin{equation}\label{mv3a}
e^{2\gamma\lambda_+(r)}=e^{2\gamma r/m_a}\biggl[1-{2\gamma\over\alpha m_a} e^{-\alpha r}+{1\over 2}{2\gamma\over\alpha m_a}\left({2\gamma\over\alpha m_a}-1\right)~e^{-2\alpha r}+\dots\biggr] , \end{equation}
and
\begin{eqnarray}\label{mv3b}
& &e^{2\gamma\lambda_-(r)}={e^{-{2\gamma r\over (b-m_p)}}\over (m_p-b)^{{2\gamma \over\alpha(m_p-b)}}}
\biggl[1+{2\gamma \over\alpha(m_p-b)}~{e^{-\alpha(r-r_p)}\over m_p-b}\nonumber\\
& & \quad \quad \quad+
{1\over 2}{2\gamma \over\alpha(m_p-b)}\left({2\gamma \over\alpha(m_p-b)}-1\right)~{e^{-2\alpha(r-r_p)}\over (m_p-b)^2}+\dots\biggr] .\end{eqnarray}
\end{subequations}
The effective potential for
the incoming comet can  be written as
\begin{subequations}
\begin{equation}\label{mv4a}
V_{eff}^+(r)=
\left[-\frac{Gm_1m_a}{r}+{l_{\theta}^2\over 2 m_a}\frac{1}{r^2}\right]
e^{2\gamma r/m_a}+W_1(\gamma,\alpha,r) ,\end{equation}
and for the outgoing comet as
\begin{equation}\label{mv4b}
V_{eff}^-(r)=
\left[-\frac{Gm_1m_a}{r}+{l_{\theta}^2\over 2 m_a}\frac{1}{r^2}\right]
{e^{{2\gamma r\over (m_p-b)}}\over (m_p-b)^{{2\gamma \over\alpha(m_p-b)}}}
+W_2(\gamma,\alpha,r)\ ,\end{equation}
\end{subequations}
where $W_1$ and $W_2$ are given in the appendix A. \\\\
\noindent
We will use  the data corresponding to the sun mass ($1.9891\times 10^{30}Kg$) and the  Halley comet ( Cevolani et al 1987,   Brandy 1982, Jewitt  2002)
$$m_c\approx 2.3\times 10^{14}Kg,\quad r_p\approx 0.6~au,\quad r_a\approx 35~au, \quad l_{\theta}\approx 10.83\times 10^{29}Kg\cdot m^2/s, \eqno(34)$$
with a mass lost of about $\quad \delta m\approx 2.8\times10^{11}Kg$ per cycle of oscillation. Although, the behavior of Halley comet seem to be chaotic (Chirikov and Vecheslavov  1989), but we will neglect this fine detail here.  Now,  the parameters $\alpha$ and $"b"$ appearing on the mass lost model, Eq. (\ref{mv1}), are determined by the chosen mass lost of the comet during the approaching to the sun and during the moving away from  the sun (we have assumed the same mass lost in each half of the cycle of oscillation of the comet around the sun). 
Using Eq. (\ref{mv4a}) and Eq. (\ref{mv4b}) in the expression  (\ref{cm13}), the trajectories can be calculated in the spaces ($r,v$) . Fig. 1 shows these
trajectories using  $\delta m=2\times 10^{10}Kg$ (or $\delta m/m=0.0087\%$) for  $\gamma=0$ and (continuos line), and for $\gamma=-3~Kg/m$ (dashed line), starting both cases from the same aphelion distance. As one can see on the minimum,  dissipation causes to reduce a little bit the velocity of the comet , and the antidissipation increases the comet velocity, reaching a further away aphelion point.  Also, when only mass lost is considered ($\gamma=0$) the comet returns to aphelion point a little further away from the initial one
during the cycle of oscillation. Something related with this effect is the change of period as a function of mass lost ($\gamma=0$). This can be see on Fig. 2, where the period is calculated starting always from 
the same aphelion point ($r_a$). Note that with a mass lost of the order $2.8\times10^{11}Kg$ (Halley comet), which correspond to $\delta m/m=.12\%$,  the comet is well within 75 years period. The variation of the ratio of the change of aphelion distance as a function of mass lost ($\gamma=0$) is shown on Fig.3.  On Fig. 4, the mass lost rate is kept fixed to $\delta m/m=0.0087\%$, and the variation of the period of the comet is calculated as a function of the dissipative-antidissipative parameter $\gamma<0$ (using $|\gamma|$ for convenience). As one can see, antidissipation always wins to dissipation, bringing about the increasing of the period as a function of this parameter. The reason seems to be that the  antidissipation acts on the comet when this ones is lighter than when dissipation was acting (dissipation acts when the comet approaches to the sun, meanwhile antidissipation acts when the comet goes away from the sun). Since the period of Halley comets has not changed much during many turns, we can assume that the parameter $\gamma$ must  vary in the interval $(-0.01,0] Kg/m$.
 Finally, Fig. 5 shows the variation, during a cycle of oscillation, of the ratio of the new aphelion ($r_a'$) to old aphelion ($r_a$) as a function of the parameter $\gamma$.\\\\
\noindent

\begin{figure}[H]
 \begin{center}
 \includegraphics[width=8.5cm,height=6cm]{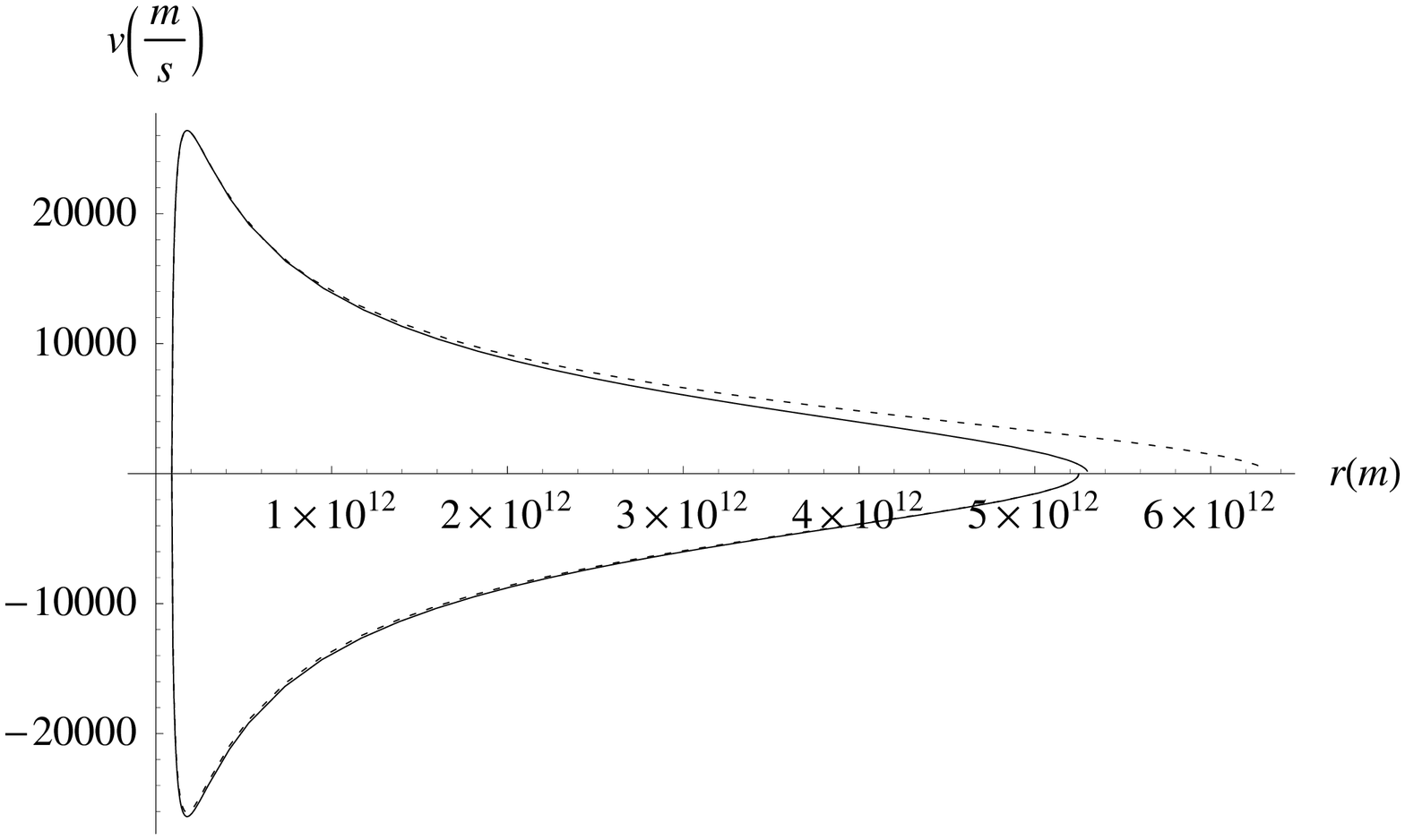}
 \end{center}
 \caption{ Trajectories in the ($r,v$) space with $\delta m/m=0.009$.}
\end{figure}

\begin{figure}[H]
 \begin{center}
 \includegraphics[width=10.5cm,height=8cm]{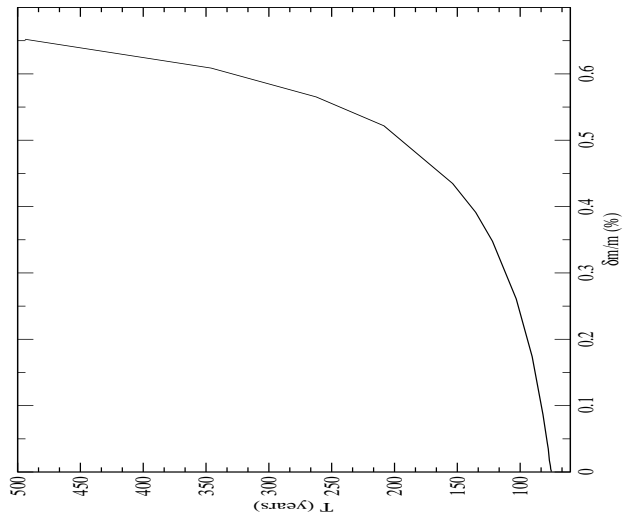}
 \end{center}
 \caption{ Period of the comet as a function of the mass lost ratio.}
\end{figure}

\begin{figure}[H]
 \begin{center}
 \includegraphics[width=8.5cm,height=6cm]{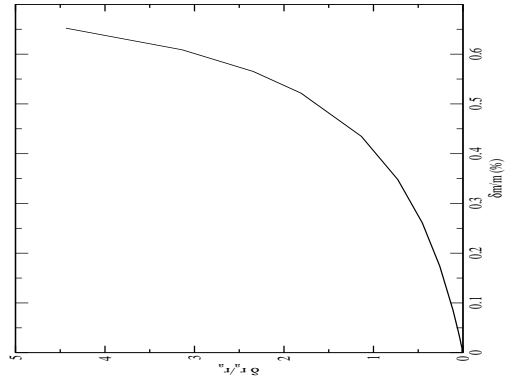}
 \end{center}
 \caption{ Ratio of aphelion distance change as a function of the mass lost rate.}
\end{figure}

\begin{figure}[H]
 \begin{center}
 \includegraphics[width=8.5cm,height=6cm]{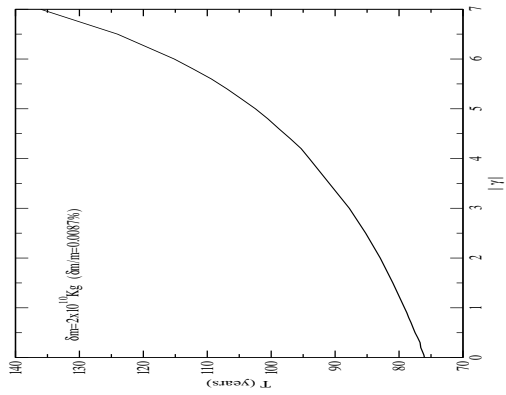}
 \end{center}
 \caption{Period of the comet as a function of the parameter $\gamma$.}
\end{figure}

\begin{figure}[H]
 \begin{center}
 \includegraphics[width=8.5cm,height=6cm]{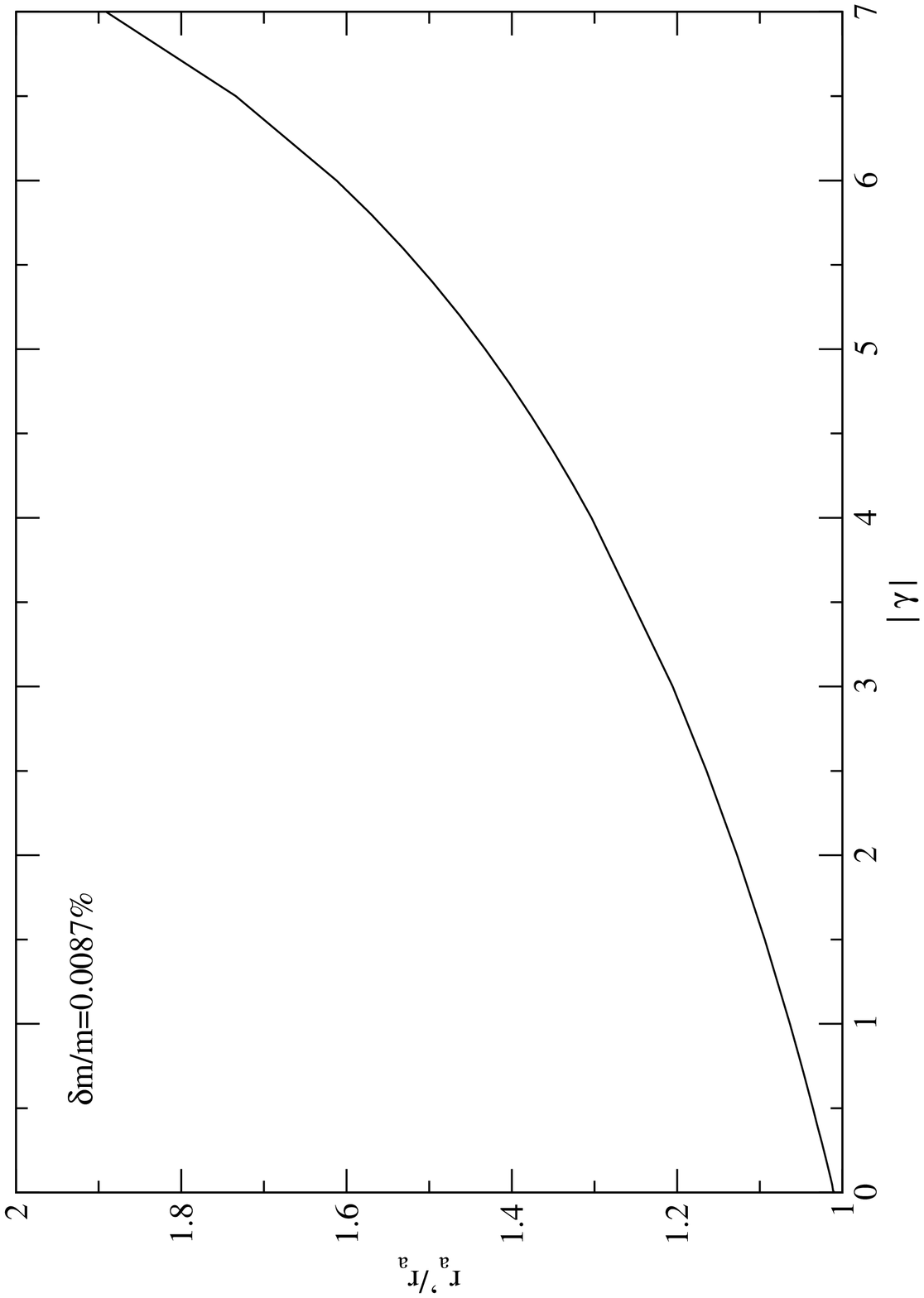}
 \end{center}
 \caption{Ratio of the aphelion increasing as a function of the parameter $\gamma$.}
\end{figure}
\section{\bf Conclusions and comments}
\noindent
We have shown that the proposed modified Newton equation for mass variation systems has some problems. Therefore, we have considered that it is better to keep Newton's equations of motion for mass variable systems to have a consistent approach to these problems. Having this in mind, the Lagrangian, Hamiltonian and a constant of motion of the gravitational attraction of two bodies were given when one of the bodies has variable mass and  the dissipative-antidissipative effect of the solar wind is considered. By choosing  the reference system in the
massive body, the system of equations is  reduce  to 1-D problem. Then, the constant of
motion, Lagrangian and Hamiltonian were obtained consistently. A model for comet-mass-variation was given, and with this model, a study was made of  the variation of the period of one cycle of oscillation of the comet when there are mass variation and  dissipation-antidissipation. When mass variation is only considered, the comet trajectory is moving away from the sun,  the mass lost is reduced as the comet is farther away (according to our model), and the period of oscillations becomes bigger. When dissipation-antidissipation is added, this former effect becomes higher as the parameter $\gamma$ becomes higher. \\\\
\newpage
\section{\bf Appendix A}
\noindent
Expression for $W_1$ and $W_2$:\\
\begin{eqnarray*}
& &W_1=\frac{Gm_{2-}^2}{m_{2+}^o}\Biggl\{
-\frac{p(p-1)e^{(-4+p)\alpha r}}{2r}+\alpha p E_i(\alpha pr)-2\alpha p(p-1)E_i\bigl((-4+p)\alpha r\bigr)\\
& &\quad+\frac{\alpha p^2(p-1)}{2}E_i\bigl((-4+p)\alpha r\bigr)+
\frac{p(p-1)}{r}\left[e^{(p-3)\alpha r}+3\alpha(1-p)r E_i\bigl((p-3)\alpha r\bigr)\right]\\
& &\quad+\frac{p(p+3)}{2}\left[-\frac{e^{(p-2)\alpha r}}{r}+\alpha(p-2)E_i\bigl((p-2)\alpha r\bigr)\right]\\
& &\quad+\frac{p+2}{r}\left[e^{(p-2)\alpha r}+\alpha(p-1)rE_i\bigl((p-1)\alpha r\bigr)\right]\Biggr\}\\
& &\quad+\frac{l_{\theta}^2}{2m_{2+}^2r^2}\Biggl\{
\frac{p(p-1)}{2}e^{(p-2)\alpha r}-pe^{(p-1)\alpha r}-\alpha p(p-1)e^{(p-2)\alpha r}+\frac{\alpha p(p-1)}{2} e^{p\alpha r}\\
& & \quad+\frac{\alpha^2p(p-1)r}{2}e^{(p-2)\alpha r}+p\alpha r e^{(p-1)\alpha r}-p^2\alpha r e^{(p-1)\alpha r}
-p^2\alpha^2 r^2 E_i\bigl(p\alpha r\bigr)\\
& &\quad-\frac{\alpha^2(p-2)^2p(p-1)r^2}{2}E_i\bigl((p-2)\alpha r\bigr)+p\alpha^2r^2 E_i\bigl((p-1)\alpha r\bigr)\\
& & \quad-2\alpha^2p^2r^2E_i\bigl((p-1)\alpha r\bigr)+p^3\alpha^2 r^2E_i\bigl((p-1)\alpha r\bigr)\Biggr\}
\end{eqnarray*}
$$\eqno(A1)$$
where $m_a$ is the mass of the body at the aphelion, and  we have made the definitions
$$p=\frac{2\gamma}{\alpha m_a}\eqno(A2)$$
and the function $E_i$ is the exponential integral, 
$$E_i(z)=\int_{-z}^{\infty}\frac{e^{-t}}{t}dt\eqno(A3)$$
\begin{eqnarray*}
& &W_2=\frac{Gm_{2-}^2}{m_{2+}^o}\Biggl\{
\frac{e^{(q-2)\alpha r}}{r}\left[1+\frac{q(q-1)}{2(m_p+\alpha q)}e^{2q\alpha r}+\frac{2q}{m_p+\alpha q}e^{q\alpha r}\right]\\
& &\quad+q\alpha E_i\bigl(q\alpha r\bigr)
-\frac{q(q-1)e^{2q\alpha r}}{(m_p+\alpha q)^2r}\left[e^{(q-3)\alpha r}-\alpha(q-3)r E_i\bigl((q-3)\alpha r\bigr)\right]\\
& &\quad +\frac{qe^{q\alpha r}}{(m_p+\alpha q) r}\left[e^{(q-3)\alpha r}-\alpha(q-3)r E_i\bigl((q-3)\alpha r\bigr)\right]
-2\alpha E_i\bigl((q-2)\alpha r\bigr)\\
& &\quad+\alpha q E_i\bigl((q-2)\alpha r\bigr)-\frac{q(q-1)\alpha e^{2q\alpha r}}{(m_p+\alpha q)^2}E_i\bigl((q-2)\alpha r\bigr)\\
& &\quad+\frac{q^2(q-1)\alpha e^{2q\alpha r}}{2(m_p+\alpha q)^2}E_i\bigl((q-2)\alpha r\bigr)-
\frac{4\alpha e^{q\alpha r}}{m_p+\alpha q}E_i\bigl((q-2)\alpha r\bigr)\\
& &\quad+\frac{2q^2\alpha e^{q\alpha r}}{(m_p+\alpha q)r}E_i\bigl((q-2)\alpha r\bigr)+
\frac{2}{r}\left[e^{(q-1)\alpha r}-(q-1)\alpha r E_i\bigl((q-1)\alpha r\bigr)\right]\\
& &\quad+\frac{qe^{q\alpha r}}{(m_p+\alpha q)r}\left[e^{(q-1)\alpha r}-(q-1)\alpha r E_i\bigl((q-1)\alpha r\bigr)\right]\Biggr\}\\
& &\quad+\frac{l_{\theta}^2}{2m_{2+}^2(m_p+\alpha q)^q}\Biggl\{
-\frac{q\alpha e^{q\alpha r}}{r}+q^2\alpha^2 E_i\bigl(q\alpha r\bigr)\\
& &+\frac{q(q-1)e^{(3q-2)\alpha r}}{2(m_p+\alpha q)^2r^2}\left[-1+2\alpha r-q\alpha r+(2-q)^2\alpha^2r^2 e^{(2-q)\alpha r} 
E_i\bigl((q-2)\alpha r\bigr)\right]\\
& & -\frac{qe^{(2q-1)\alpha r}}{(m_p+\alpha q)r^2}
\left[-1+\alpha r+q\alpha r+(q-1)^2\alpha^2r^2 e^{(1-q)\alpha r} E_i\bigl((q-1)\alpha r\bigr)\right]\Biggr\}
\end{eqnarray*}
$$\eqno(A4)$$
where $m_p$ is the mass of the body at the perihelion, and we have made the definition
$$q=\frac{2\gamma}{\alpha(m_p-b)}\eqno(A5)$$ 
\newpage


\newpage
\begin{thebibliography}{10}

Bekov A.A., 1989, 
\newblock{ Astron. Zh.,}  66, 135\\

Berkovich L.M., 1981, 
\newblock{Celestial Mechanics,}   24 ,407\\

Bethe H.A., 1986,  
\newblock{ Phys. Rev. Lett.,} 56, 1305\\

 Brandy J.L., 1982, 
\newblock{ J. Brit. Astron. Assoc.,} 92, no. 5, 209\\

Cevolani G., Bortolotti G. and  Hajduk A., 1987,   
\newblock{ IL Nuo. Cim. C,}  10, no.5, 587\\

Chirikov B.V. and Vecheslavov V.V., 1989, 
\newblock{ Astron. Astrophys.,}  221,  146\\

 Commins E.D. and Bucksbaum P.H., 1983, 
\newblock { Weak Interactions of Leptons  and Quarks}, 
\newblock{Cambridge University Press }\\
%
Daly P.W.,  1989, 
\newblock{ Astron. Astrophys.,} 226, 318\\
%
Goldstein H., 1950, 
\newblock{ Classical Mechanics},
\newblock{ Addison-Wesley, M.A.}\\

Gylden, H., 1884,
\newblock {Astron. Nachr.,}  109, no. 2593,1\\

Helhl F.W.,  Kiefer C. and Metzler R.J.K., 1998, 
\newblock{ Black Holes: Theory  and Observation}, 
\newblock{Springer-Verlag}\\
%
 Jeans J.H., 1924, 
\newblock {MNRAS,}   85, no. 1, 2\\

Jewitt D.C., 2002,
\newblock{ Astron. Jour.,}  123, 1039\\ 

John F., 1974,
\newblock{ Partial Differential Equations}, 
\newblock{Springer-Verlag, New York}\\

Kobussen J.A., 1979,
\newblock{Acta Phys. Austr.}  51,193\\
Leubner C., 1981,
\newblock{ Phys. Lett. A,}  86, 2\\

L\'opez G.,  Barrera L.A.,  Garibo Y., Hern\'andez H.,  Salazar J.C., 2004,
\newblock {Int. Jour. Theo. Phys.,}  43, no. 10,  1\\

L\'opez G., 1999,
\newblock{ Partial Differential Equations of First Order and
Their Applications to Physics}, 
\newblock{World Scientific}\\

 L\'opez G., 1996, 
\newblock{ Ann. of  Phys.,} 251, no. 2 , 372\\
%
L\'opez G, 2007, 
\newblock{ Int. Jour. Theo. Phys.,} 46, no. 4, 806\\
%
Lovett E.O., 1902, 
\newblock{ Astron. Nachr.}, 158, no. 3790, 337\\
%
 Meshcherskii I.V., 1893,
\newblock{ Astron. Nachr.,}  132, no. 3153, 93\\

Meshcherskii I.V., 1902, 
\newblock{ Astron. Nachr.,}  159, no. 3807, 229\\
%
 M{\o}ller C., 1952
 \newblock{\it Theory of Relativity}, 
 \newblock{Oxford University Press.}\\
 %
Plastino A.R. and Muzzio J.C., 1992, 
\newblock{ Cel. Mech. Dyn. Ast. }, 53, 227\\
%
Prieto C. and  Docobo J.A.,  1997,
\newblock{ Astron. Astrophys.,}  318, 657\\

Serimaa O.T., Javanainen J., and  Varr\'o S., 1986, 
\newblock{ Phys. Rev. A,}  33, 2913\\ 

Spivak M.,2010,
\newblock{Physics for Mathematicians, Mechanics I}
\newblock{Publish or Perish Inc., chapter 3}\\

Sommerfeld A., 1964,
\newblock{ Lectures on Theoretical Physics}, Vol. I, 
\newblock{Academic Press}\\
%
Zagorodny A.G., Schram P.P.J.M., and  Trigger S.A., 2000, 
\newblock{ Phys. Rev. Lett.,}   84, 3594\\
%

 %
\end{thebibliography}
\end{document}